# Comment on

## *Unusual violation of the Wiedemann–Franz law at ultralow temperatures in topological compensated semimetals*

## by Yi-Yan Wang et al.

Kamran Behnia[1], Shiyan Li,[2,3] Johnpierre Paglione,[4,5] & Louis Taillefer[5,6]

1) Laboratoire de Physique et d'Etude de Matériaux (CNRS), ESPCI Paris, Université PSL, 75005 Paris, France
2) State Key Laboratory of Surface Physics and Department of Physics, Fudan University, Shanghai 200438, China
3) Shanghai Research Center for Quantum Sciences, Shanghai 201315, China
4) Maryland Quantum Materials Center and Department of Physics, University of Maryland, College Park, Maryland 20742, USA
5) Canadian Institute for Advanced Research, Toronto, Ontario, Canada
6) Institut quantique, Département de physique and RQMP, Université de Sherbrooke, Sherbrooke, Québec, Canada

Kamran.behnia@espci.fr

shiyan_li@fudan.edu.cn

paglione@umd.edu

louis.taillefer@usherbrooke.ca

*Abstract-* Recently, Wang et al. [1] reported on an `unusual' violation of Wiedemann-Franz law in three semimetals. We compare their observations to our observations in a variety of systems, where the apparent WF law violations in the same temperature range arise as a consequence of electron-phonon decoupling. Given the empirical similarity of their data with these cases, the most plausible explanation for the reported violation is an experimental artefact.

Measuring thermal ($\kappa$) and electrical ($\sigma$) conductivities down to sub-Kelvin temperatures, Wang *et al.* [1] reported an "unusual" violation of the Wiedemann–Franz (WF) in the semimetal materials $TaAs_2$, $NbAs_2$ and $NdSb$. Their conclusion was derived from the observation of a rapid decrease of thermal conductivities of all samples below approximately 200 mK, with $\kappa(T)$ following a strong ~$T^4$ decrease below this temperature. Without a concomitant decrease in the electrical conductivities, which remain mostly temperature-independent through this range, the strong deviation between thermal and electrical conductivities led Wang *et al.* to conclude that there is a strong and `unusual' violation of WF law in these topological semimetals. Here, we compare these observations to very similar observations in a diverse set of systems, where the apparent WF law violations in the same temperature range have been confirmed to arise due to electron-phonon decoupling, and not an intrinsic violation of this robust law.

The WF law [2] links the ratio of the amplitudes of electronic thermal conductivity $\kappa_e$ and electrical conductivity $\sigma$ of metals to fundamental constants: $\kappa_e/T = (\pi^2/3) (k_B/e)^2 \sigma$. Numerous experiments have found this law to be extremely robust in the zero-temperature limit (which is why it is routinely used to test the accuracy of an experimental set-up designed to measure the thermal conductivity of metallic samples), and its true violation would imply a marked deviation from expectations of the quasiparticle model of electron behavior in metals, the so-called Fermi liquid theory.

Nearly 25 years ago, a rapidly vanishing thermal conductivity like the one reported in [1] was observed in cuprate superconductors [3]. Initially, it was observed in the context of verifying the validity of the WF law in an (optimally doped) electron-doped cuprate [3]. But the subsequent observation of a similar downturn in the thermal conductivity of a heavily overdoped non-superconducting cuprate [4], established as a Fermi liquid [4], suggested that an experimental artifact was behind the apparent violations. Indeed, a careful examination found that such downturns will occur in milliKelvin thermal conductivity measurements due to the decoupling between electronic and phononic thermal reservoirs at low temperatures [5].

In a standard one-heater, two-thermometer steady-state measurement, thermal conductivity is measured by quantifying the local temperature gradient imposed by the presence of a heat current. Temperature sensors (usually resistors) coupled to the sample are used to monitor the local temperature at two contacts to the sample, and the thermal conductance is derived from the ratio of heat current to temperature difference. This measurement routinely assumes that the measured temperatures are indeed the intrinsic temperatures of the heat carriers in the sample, typically dominated by electrons at the lowest temperatures. However, in the sub-Kelvin temperature range, the Kapitza conductance between the electronic and phononic reservoirs (which roughly follows a quartic $T^4$ power law [5]) can become much smaller than the purely electronic thermal conductance. In this case, if the thermal path between the sensors and the electron reservoir is more resistive than the one between the sensor and the phonon reservoir, the measured temperature will not reflect the true temperature of the electronic system, but rather the phonon bath temperature. In analogy, the path between heater and thermometer is governed by the same constraint, such that thermal contacts that are too electrically resistive – for either thermometers, thermal ground or heater contacts – will limit thermal contact to the electron temperature, resulting in the observation of a pathological downturn at a characteristic temperature that reflects this decoupling [5].

Subsequent to the work of Smith et al [5], a consensus emerged that the apparent zero-temperature WF law violation measured in the cuprates is indeed caused by such electron-phonon decoupling and not by any exotic physics, and is understood to stem from the experimental (and material-dependent) limitation of making a robust thermal path between the electron reservoir and the temperature sensor. More recently, this effect has been confirmed in several other otherwise unrelated materials, all of which have found an unexpected low temperature decrease in the measured electronic thermal conductivity. Fig. 1 shows a comparison between the temperature dependence of the reported thermal conductivity divided by temperature, $\kappa/T$, in NdSb [1] and in four other metals. In PCCO [3], LSCO [4], NCCO [6] and UTe$_2$ [7], the observed downturn has been diagnosed to be a consequence of the decoupling between electrons and phonons. In three cases (PCCO [3], NCCO [6] and UTe$_2$ [7]), the temperature dependence of the data has been quantitively accounted for by using an expression derived from a simple model in [5]. The same observation was also made in the case of the heavy-fermion metal CeCoIn$_5$ (see Fig. S8 in the supplement of [8]), where an increase in contact resistance by a factor of 100 was explicitly shown to by well explained by electron-phonon decoupling. We also note that it is not only electrons that can decouple from the phonon bath, but also magnons and more exotic excitations such as spinons (spinon-phonon decoupling is unavoidable, as indeed observed in copper benzoate -- see Fig. 1f of Ref. [9]).

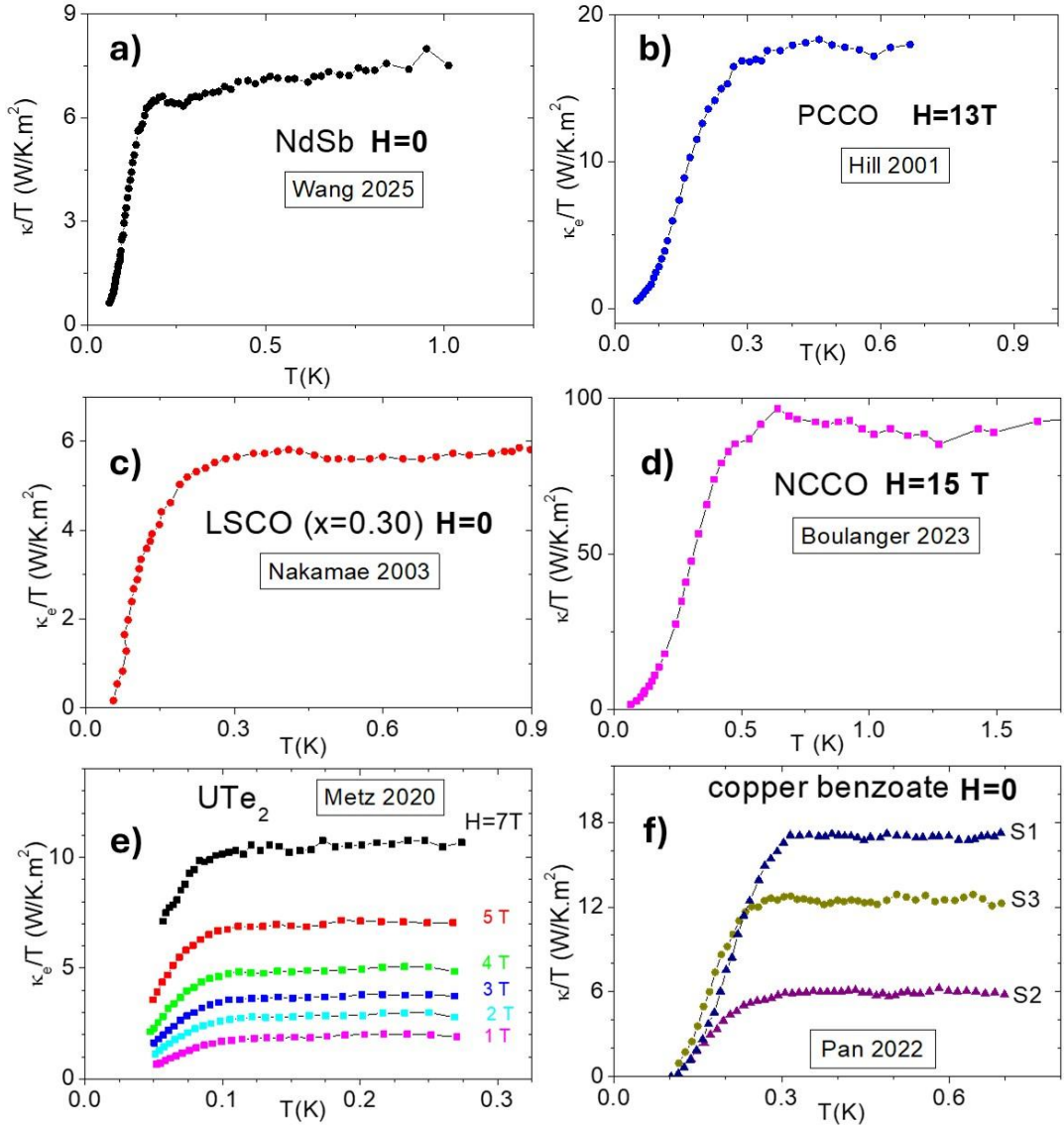

**Figure 1**- *Comparison of the temperature dependence of the apparent thermal conductivity in: (a) NdSb [1], (b) optimally-doped $Pr_{2-x}Ce_xCuO_{7-\delta}$ (PCCO) [3], (c) over-doped $La_{1.7}Sr_xCuO_4$ (LSCO) [4], (d) optimally-doped $Nd_{2-x}Ce_xCuO_4$ (NCCO) [5], (e) $UTe_2$ [7] and (f) $Cu(C_6H_5COO)_2 \cdot 3H_2O$ (copper benzoate) [9]. Note that in panels b, c and e, the plotted quantity is the electronic component of the thermal conductivity, $\kappa_e$. In panel f, the fermionic carriers of heat are not electrons, but magnetic excitations. In panels b to f, the observed downturn occurs at a temperature below which the measured temperature difference is different*

*from the temperature difference inside the fermionic thermal bath. The latter is decoupled from the phonon bath, as explained in the text.*

Given the striking empirical similarity of the low-temperature behavior of the thermal conductivity of TaAs$_2$, NbAs$_2$, and NdSb with all of the cases noted above, in particular the similar temperature dependence and range of temperature where the phenomena occur, the most plausible explanation for the observed WF law violation reported in [1] is electron-phonon decoupling, and not a true, intrinsic violation of this robust law.